# HIV/AIDS in South Africa: the beginning of the end?

Brian G. Williams,* Eleanor Gouws† and John Hargrove*

\* South African Centre for Epidemiological Modelling and Analysis, Stellenbosch, South Africa
† Joint United Nations Programme on HIV/AIDS (UNAIDS), Geneva, Switzerland

Correspondence to BrianGerardWilliams@gmail.com

**Abstract**

In several countries in southern Africa, including South Africa, the prevalence of HIV remains stubbornly high in spite of considerable efforts to reduce transmission and to provide anti-retroviral therapy (ART). It is important to know the extent to which the high prevalence of HIV reflects the increasing number of people on ART in which case the prevalence of those not on ART may be falling. Unfortunately, direct measures of the proportion of HIV-positive people who are on ART are lacking in most countries and we need to use dynamical models to estimate the impact of ART on the prevalence of HIV. In this paper we show that the current level of ART provision in South Africa has probably reduced the prevalence of HIV among those not on ART by 1.9 million, averted 259 thousand new infections and 428 thousand deaths.

## Introduction

South Africa has one of the worst epidemics of HIV in the world,[1] but has the best surveillance system[2-5] and now one of the highest levels of ART provision.[1]

In some countries, including most notably Zimbabwe,[6] HIV prevalence is falling. While it must be the case that people's behaviour has changed, the reasons for this are not immediately apparent. In South Africa, on the other hand, the prevalence of HIV among women attending ante-natal clinics has been stable for several years with no sign of a decline.[7] Here we ask, in relation to South Africa in particular, what is the likely impact of ART on the incidence, prevalence and mortality of HIV in South Africa?

## Methods

In the absence of direct empirical data we rely on modelling the impact of ART on the HIV epidemic. We fitted a previously published model[8] to trend data on HIV prevalence from the annual ante-natal clinic (ANC) surveys, scaled to reflect the rates in the overall adult population aged 15 years or more.[8] The model includes four stages of HIV-infection, with equal rates of progression between them, which gives a good approximation to a Weibull distribution with a mean survival time of 11.5 years and a shape parameter of 2.25.[9] We use ART coverage data for South Africa reported by WHO and UNAIDS.[1] Given that the mean $CD4^+$ cell count in young adults in South Africa, immediately after seroconversion, is about $800/\mu L$,[10] Stage 4 corresponds to people with a $CD4^+$ cell count between 0 and $200/\mu L$. We assume that everyone starts ART in Stage 4.[11] We also assume that people starting ART will extend their life expectancy by a factor of 4 depending on the stage they are in when they start treatment.[12] The median life-expectancy of those entering Stage 4 will therefore increase from 2.9 years to 11.5 years. This is consistent with recent data for people starting ART at very low $CD4^+$ cell counts among whom there is 80% to 90% survival at 5 years even among those starting ART at a $CD4^+$ cell count of $15/\mu L$.[13]

## Results

Figure 1A shows the model fitted to the current estimates of HIV prevalence (blue dots) and the projected HIV incidence (green line), prevalence (blue line) and deaths (black line) if we assume that ART had never been provided and people had not changed their behaviour.

In Figure 1B, the proportion of people in ART (pink line) increases logistically with the parameters chosen to give the best fit to the ART data (pink dots). We allow the coverage of ART to increase until 2020, as shown in the Figure. Projections into the future will depend on assumptions about the extent to which ART coverage continues to increase and on the life-expectancy after starting ART.

The modelled prevalence of HIV, including those on and off ART, matches the measured data quite precisely in both cases. However, Figure 1B suggests that there has been a significant decline in the prevalence of HIV among those not on ART (red line) while those on ART (pink line) will live for another ten years and possibly more. Indeed the model suggests that the ART programme in South Africa has already reduced the prevalence of HIV among those not on ART by 1.9 million, has averted 259 thousand new infections and 428 thousand deaths. Maintaining and slightly expanding the current level of treatment up to 2010 will reduce the prevalence of HIV in those not on treatment by a further 1.4 million, avert and additional 1.3 million new infections and 1.3 million deaths.

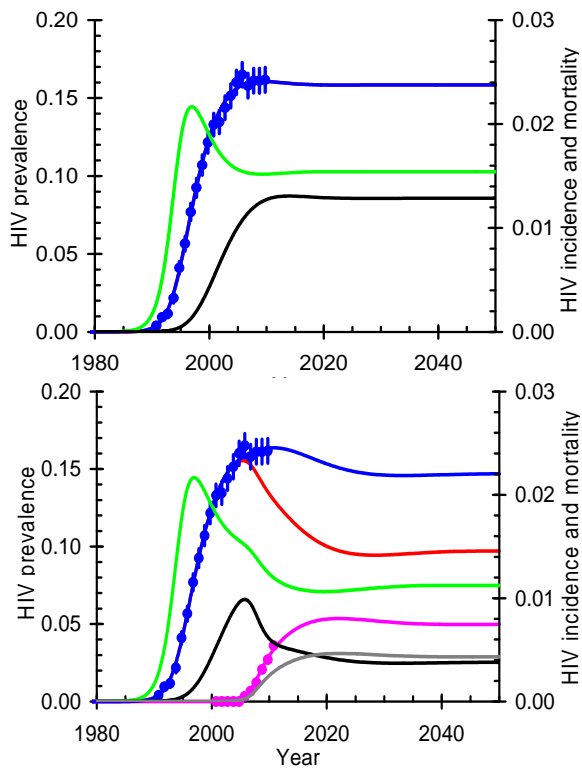

Figure 1. A: Adult prevalence (blue), annual incidence (green) and mortality (black) fitted to the trend in the prevalence of HIV in South Africa (blue dots and error bars) assuming that ART had not been provided. B: As in A but with the prevalence of HIV in people not on ART (red), on ART (pink), and the annual mortality of people not on ART (black) and on ART (grey). Proportion of people receiving ART (pink dots).

It is important to note that while starting people in South Africa on ART when their $CD4^+$ cell count is below 200/μL reduces the steady mortality by a factor of 3.3, from 1.3% p.a. to 0.4% p.a., it only reduces incidence by a factor of 1.5, from 1.6% p.a. to 1.1% p.a. To further reduce incidence as well as mortality it will be necessary to start ART at higher $CD4^+$ cell counts and to eliminate HIV transmission people need to start ART within one year of sero-conversion.[8]

## Discussion

The roll-out of ART in South Africa has had, and will continue to have, a substantial impact on the epidemic of HIV, reducing the death-toll from AIDS by 1.7 million over the next 20 years. However, the ultimate arbiter of science is not the authority of scientists or the quality of theoretical models; rather it lies in actual data, carefully measured, collected and analyzed. What models can do is tell us what more we need to know. In particular, it would be of great value if, in the next ANC survey in South Africa, blood samples of HIV positive cases were both used to estimate HIV incidence[14] and tested for ART residues. This would allow us to make better estimates of the actual and predicted impact of the ART roll-out on prevalence and incidence, and give us much greater confidence in what seems to us to be a significant achievement in slowing down the epidemic of HIV. It would also help to improve our model predictions and inform us on the best way to bring the epidemic to an end.

## References

1. UNAIDS. Report on the Global AIDS Epidemic. 2010; Available from: http://www.unaids.org/documents/20101123_GlobalReport_em.pdf
2. Ingram P. Aids figures may be too optimistic. Focus (Journal of the Helen Suzman Foundation) 2000; Available from: www.hsf.org.za/resource-centre/focus/issues-20-11/ issue-18-second-quarter-2000
3. Swanevelder JP, Kustner HG, van Middelkoop A. The South African HIV epidemic, reflected by nine provincial epidemics, 1990-1996. S Afr Med J. 1998; **88**(10): 1320-5.
4. Kustner HG, Swanevelder JP, van Middelkoop A. National HIV surveillance in South Africa, 1993-1995. S Afr Med J. 1998; **88**(10): 1316-20.
5. Kustner HG, Swanevelder JP, Van Middelkoop A. National HIV surveillance-South Africa, 1990-1992. S Afr Med J. 1994; **84**(4): 195-200.
6. Hargrove JW, Humphrey JH, Mahomva A, Williams BG, Chidawanyika H, Mutasa K, *et al.* Declining HIV prevalence and incidence in perinatal women in Harare, Zimbabwe. Epidemics. 2011; **3**: 88-94.
7. Anonymous. National Antenatal Sentinel HIV and Syphilis Prevalence Survey in South Africa, 2009: Department of Health; 2010.
8. Granich RM, Gilks CF, Dye C, De Cock KM, Williams BG. Universal voluntary HIV testing with immediate antiretroviral therapy as a strategy for elimination of HIV transmission: a mathematical model. Lancet. 2009; **373**(9657): 48-57.
9. Williams BG, Granich R, Chauhan LS, Dharmshaktu NS, Dye C. The impact of HIV/AIDS on the control of tuberculosis in India. Proc Nat Acad Sc USA. 2005; **102**(27): 9619-24.
10. Williams BG, Korenromp EL, Gouws E, Schmid GP, Auvert B, Dye C. HIV Infection, antiretroviral therapy, and CD4+ cell count distributions in African populations. J Infect Dis. 2006; **194**(10): 1450-8.
11. World Health Organization. Towards Universal Access: Scaling Up Priority HIV/ AIDS Interventions in the Health Sector. Progress report. 2010.
12. Williams BG. Determinants of sexual transmission of HV: implications for control. arXiv; 2011 http://arxiv.org/abs/ 1108.4715.
13. Walker AS, Ford D, Gilks CF, Munderi P, Ssali F, Reid A, *et al.* Daily co-trimoxazole prophylaxis in severely immunosuppressed HIV-infected adults in Africa started on combination antiretroviral therapy: an observational analysis of the DART cohort. Lancet. 2010; **375**(9722): 1278-86.
14. UNAIDS/WHO Working Group on Global HIV/AIDS and STI Surveillance. When and how to use assays for recent infection to estimate HIV incidence at a population level. Geneva: UNAIDS/WHO; 2011 http://www.who.int/diagnostics_laboratory/hiv_incidence_may13_final.pdf.